# Pair weakening in mixed rare-earth nickel borocarbides ($R,R'$)Ni$_2$B$_2$C

J Freudenberger[1], K-H Müller[1], G Fuchs[1], M Wolf[1], K Nenkov[1,2], L Schultz[1]

E-mail: J.Freudenberger@ifw-dresden.de

[1] IFW Dresden, Leibniz-Institut für Festkörper- und Werkstoffforschung Dresden e.V.,
P.O. Box 270116, D-01171 Dresden, Germany
[2] Institute of Solid State Physics, BAS, Sofia, Bulgaria

**Abstract.** In this study pseudoquaternary rare-earth nickel borocarbide superconductors $R_xR'_{1-x}$Ni$_2$B$_2$C have been investigated predominantly in the diluted limit $x \ll 1$ or $(1-x) \ll 1$. In all of these materials structural disorder results in a reduction of the superconducting transition temperature $T_c$. Depending on the selection of the rare earth elements $R$ and $R'$ this disorder induced deterioration of superconductivity is combined with magnetic pair breaking of Abrikosov-Gor'kov type or pair breaking of non-magnetic impurities in antiferromagnetic superconductors (Morozov-type of pair breaking).

## 1. Introduction

The superconducting transition temperature $T_c$ of superconductors can be changed by chemical modification, in particular by doping the material with impurities. Since in most of the known superconductors the Cooper pairs are spin singlets which rely on the presence of time-reversal symmetry, the influence of magnetic impurities on superconductivity is very different from that of non-magnetic ones [1].

*1.1. Magnetic pair-breaking*

Magnetic impurities are strong pair breakers because they lift the time-reversal symmetry. Therefore the superconducting transition temperature $T_c$ is reduced if magnetic impurity ions are introduced in the host matrix of a conventional superconductor. This reduction of $T_c$ can be very large. Thus one percent of Gd in superconducting Lanthanum reduces $T_c$ from about 6 $K$ to zero (see Table 1) [2].

In a series of superconductors with paramagnetic impurities the variation of $T_c$ in dependence on the impurity content is represented by the expression

$$\ln\left(\frac{T_c^0}{T_c}\right) = \psi\left(\frac{N(E_F)I^2 \overline{DG}}{2T_c}\right) - \psi\left(\frac{1}{2}\right) \qquad (1)$$

of the Abrikosov Gor'kov theory [3], where $T_c^0$ is the superconducting transition temperature of the undoped parent system. $N(E_F)$ is the conduction electron density of states at the Fermi level, $I$ is a strength parameter for the exchange between conduction electrons and the localized magnetic electrons of the impurities, $\psi$ is the digamma function, $\overline{DG} = xDG[R'] + (1-x)DG[R]$ is the effective de Gennes factor, $x$ is the concentration of the magnetic impurities $R'$, and $DG[R]$ and $DG[R']$ are the corresponding de Gennes factors of $R$ (and $R'$)

$$DG[R] = (g-1)^2 J(J+1) \qquad (2)$$

of the lattice sites (here assumed to be occupied by rare-earths $R$ and $R'$, respectively), with $g$ as the Landé factor and $J$ the total angular momentum of the $R^{3+}$ Hund's-rule ground state. In a non-magnetic parent compound there will be $DG[R] = J = 0$.

Table 1 shows the reduction of $T_c$ of La substituted by 1 at.% by other rare earths $R'$. This table also contains some electronic properties of the $R'$ elements and the $R'^{3+}$ ions carrying 4f-electron based local magnetic moment in these doped materials. It can be easily shown that the reduction of the superconducting transition temperatures in Table 1, $\delta T_c$, is well reproduced by the Abrikosov Gor'kov formula (1).

**Table 1.** Electron configuration of the rare-earth metals $R'$, total angular momentum $J$, spin $S$, Landé factor $g$, de Gennes factor DG, paramagnetic moment of the trivalent ion $R'^{3+}$ and the reduction of the superconducting transition temperature, $\delta T_c$, for 1 at.% $R'$ solid solution in superconducting Lanthanum after Ref. [2].

| element | electron conf. : [Xe]-core + 5d+6s | 4f | $J$ | $S$ | $g$ | de Gennes factor $(g-1)^2 J(J+1)$ | para. mom. $\mu_B$ | $\delta T_c$ [K] |
|---|---|---|---|---|---|---|---|---|
| La | 3 | 0 | 0 | 0 | - | | 0 | 0 |
| Ce | 3 (2) | 1 (2) | 5/2 | 1/2 | 6/7 | 0.179 | 2.54 | -2.1 |
| Pr | 3 (2) | 2 (3) | 4 | 1 | 4/5 | 0.8 | 3.58 | -0.5 |
| Nd | 3 (2) | 3 (4) | 9/2 | 3/2 | 8/11 | 1.841 | 3.62 | -1.0 |
| Pm | 3 (2) | 4 (5) | 4 | 2 | 3/5 | 3.2 | 2.68 | |
| Sm | 3 (2) | 5 (6) | 5/2 | 5/2 | 2/7 | 4.464 | 0.85 | -1.2 |
| Eu | 3 (2) | 6 (7) | 0 | 3 | - | | 7.94 | -1.9 |
| Gd | 1+2 | 7 | 7/2 | 7/2 | 2 | 15.75 | 7.94 | -5.0 |
| Tb | 1+2 | 8 | 6 | 3 | 3/2 | 10.5 | 9.72 | -2.2 |
| Dy | 3 (2) | 9 (10) | 15/2 | 5/2 | 4/3 | 7.083 | 10.65 | -1.9 |
| Ho | 3 (2) | 10 (11) | 8 | 2 | 5/4 | 4.5 | 10.61 | -0.6 |
| Er | 3 (2) | 11 (12) | 15/2 | 3/2 | 6/5 | 2.55 | 9.58 | -0.4 |
| Tm | 3 (2) | 12 (13) | 6 | 1 | 7/6 | 1.167 | 7.56 | |
| Yb | 3 (2) | 13 (14) | 7/2 | 1/2 | 8/7 | 0.321 | 4.53 | -0.2 |
| Lu | 1+2 | 14 | 0 | 0 | - | | | -0.1 |

*1.2. Disorder*

According to the Anderson theorem [1] non-magnetic impurities can not break spin-singlet Cooper pairs. Therefore non-magnetic impurities can not "directly" reduce the value of $T_c$ and materials with strong non-magnetic disorder can be superconducting (so called dirty superconductors [1]). Nevertheless non-magnetic impurities can indirectly modify $T_c$ because they affect the electronic and phononic properties of the doped material through various mechanisms as e.g. by electron transfer, change of the conduction electron density $N(E_F)$, local lattice strains caused by deviations of the atomic or ionic size of the substituting atoms from that of the host atoms or ions, respectively, change of the phonon spectrum, whereby these effects are not independent of each other and are very difficult to describe quantitatively.

A typical example for electronic disorder effects is a shift of the peak of the density of states at the Fermi level, combined with broadening of that peak. On the other hand, in multiband superconductors the rate of scattering of the conduction electrons between different electron bands may increase [4,5].

Attfield et al. [6] have shown that in $(R,M)_2CuO_4$ superconductors, $T_c$ can be well described, on a phenomenological level, by assuming a random distribution of the $R^{3+}$ and $M^{2+}$ cations on the $R$ lattice site and relating $T_c$ to the variance $\sigma^2$ of cation radius at that site.

*1.3. Non-magnetic impurities in superconducting antiferromagnets*

An interesting theoretical prediction is that, similarly as in the spin-triplet paired superconductors non-magnetic impurities in an antiferromagnetic superconductor cause pair breaking [7,8,9]. It has been pointed out by Gupta [10], that such depression of superconductivity in antiferromagnetic superconductors by non-magnetic impurities may be the reason why not many antiferromagnetic superconductors with $T_c < T_N$ are known, where $T_N$ is the antiferromagnetic ordering temperature. In principle there is no reason as to why many more such materials should not exist. However, in most such cases $T_c$ may already have been suppressed, beyond observation, by non-magnetic disorder that is always present to some degree. As a consequence of this phenomenon, the value of $T_c$ of $DyNi_2B_2C$ is very sensitive to the presence of non-magnetic impurities or, more generally, to the detailed metallurgical state of the samples. Possibly for that reason the identification of superconductivity in $DyNi_2B_2C$ was delayed compared to the other borocarbide superconductors and the published experimental data on the properties of $DyNi_2B_2C$ and of Dy-rich pseudoquaternary compounds $R_xDy_{1-x}Ni_2B_2C$ (x << 1) exhibit much scatter [5,10].

*1.4. Superconducting borocarbides*

In rare earth nickel borocarbides $RNi_2B_2C$ ($R$ = rare earth, Sc, Y or La) all the previously mentioned mechanisms occur and coexist. These materials which were discovered in 1994 [11,12] have generated large interest because they have relatively high transition temperatures (up to about 17 $K$) and they show coexistence of superconductivity and long range magnetic ordering [13]. Soon after the discovery of these quaternary borocarbide superconductors a remarkable progress in the investigation of their physical properties could be asserted [14]. They have been considered as "a toy box for solid-state physicists" [15] and the study of them resulted in better understanding of superconductors in general and magnetic superconductors in particular [5,10]. A typical example is the novel concept of strongly coupled two-band superconductivity introduced to explain the anomalous temperature dependence of the upper critical field of $YNi_2B_2C$ and $LuNi_2B_2C$ [4] and now successfully used for other novel superconductors such as $MgB_2$ or (possibly) FeAs-based compounds [16].

As the different $R^{3+}$ ions in the $RNi_2B_2C$ compounds can be replaced as well as substituted by each other nearly in the whole range of concentration $x$, a large number of series of $R_xR'_{1-x}Ni_2B_2C$ compounds with very different properties can be formed and investigated.

## 2. Experimental details

Polycrystalline $R_xR'_{1-x}Ni_2B_2C$ samples were prepared by a standard arc melting technique. Powders of the elements were weighted in the stoichiometric compositions with a surplus of 10 wt.% boron to compensate the high losses of boron caused by the arc melting. The powder was pressed to pellets which were melted under argon gas on a water-cooled copper plate in an arc furnace. To get homogeneous samples, they were turned over and melted again four times. After the melting procedure the solidified samples were homogenised at 1100°C for ten days. As the actual sample composition is somewhat uncertain, the nominal sample composition is given in this paper.

Using this method single-phase samples containing $R'$ = La can not be obtained in the concentration range $0.4 < x < 0.7$. The rapid quenching single-roller melt-spinning technique was used to suppress this phase separation. For this purpose ingots of nominal composition were prepared by arc melting as above. Each ingot was heated at radio frequency in a quartz jet over the melting temperature and then pressed out by an overpressure. The melt was quickly solidified by getting into contact with the copper wheel (see Ref. [17]).

A powder X-ray diffractometer in Bragg-Brentano-geometry was used to verify the phase purity of > 95% and also the lattice structure. The X-ray diffraction experiments were performed on crushed powders using $Co_{K\alpha}$ ($\lambda$ = 0.1789 nm) radiation. The scans were taken from $2\Theta = 20°$ up to $120°$ in steps of $\Delta\Theta = 0.02°$.

Single-phase samples containing Ce, Eu or Yb could not yet be prepared for various reasons. Ce and Eu do not form the $LuNi_2B_2C$-type structure because the 4f electrons of these elements tend to stronger hybridize with the conduction electrons in intermetallic compounds, compared to the behaviour of the other $R'$ elements [18]. Yb has a very high vapour pressure resulting in high Yb losses during melting, that can not be compensated and cause off-stoichiometric or multi-phase samples.

Single crystals of $R_{0.01}R'_{0.99}Ni_2B_2C$ (with $R$ = Ho, Dy; $R'$ =Y, La, Lu) were grown from $Ni_2B$ flux. For this purpose the flux growth technique was utilized. A polycrystalline sample of $R_{0.01}R'_{0.99}Ni_2B_2C$ stoichiometric composition was placed onto a $Ni_2B$ sample in a carbon crucible. This arrangement was heated in Argon atmosphere with a rate of 300 K/h up to a temperature of 1550°C and held at this temperature for two hours. During cooling down to 1200°C with a rate of 5 K/h single crystals were grown at the $R_{0.01}R'_{0.99}Ni_2B_2C$ rich site. Below 1200°C the arrangement was cooled down to room temperature with a rate of 300 K/h. The typical size of these single crystals is 1x1x 0.2 mm$^3$.

All samples were investigated by susceptibility and resistivity measurements to determine the superconducting transition temperature $T_c$ and its width $\Delta T_c$.

## 3. Results and discussion

The X-ray patterns of the investigated polycrystalline samples (not shown here) revealed no significant fraction of phase impurities. As a result of the X-ray structure analysis all of the $R_xR'_{1-x}Ni_2B_2C$ samples reveal the $LuNi_2B_2C$-type structure (space group I/4mmm) [19] in the whole range of concentration $x$. This confirms that substituting $R$ by another $R'$ in $RNi_2B_2C$ compounds does not change their lattice symmetry [19,20].

The lattice parameters $a$ and $c$ vary linearly in this range between the values of the pure $RNi_2B_2C$ compounds. For La containing samples and intermediate concentrations, $0.4 \leq x \leq 0.7$, two patterns with the I/4mmm space group were found as described in our former work [17].

The superconducting transition curves (not shown here) reflect the phase purity, too, as they reveal sharp jumps with widths up to $\Delta T_c \approx 0.3$ K. The quantity $\Delta T_c$ was determined as $\Delta T_c = T(0.9\chi_N) - T(0.1\chi_N)$, where $\chi_N$ is the normal state susceptibility. This holds for single crystals and arc-molten polycrystals, only. Rapidly quenched samples – if superconducting – show larger values of $\Delta T_c$.

In Fig. 1 the superconducting transition temperature $T_c$ of $Ho_xR_{1-x}Ni_2B_2C$ and $Dy_xR_{1-x}Ni_2B_2C$ compounds ($R$ = Y, Lu, La) is shown in dependence on the effective de Gennes factor. The individual de Gennes factors are $DG[Ho] = 4.5$ and $DG[Dy] = 7.083$ (see Table 1 and Eq. (2)). The curves in Fig. 1 vary much in their shape ranging from the linear curve of the Ho–Y system to non-linear curves in the other cases and the even non-monotonic curve for $Dy_xLu_{1-x}Ni_2B_2C$. This complex behaviour is due to the fact that, as discussed in Sec. 1, very different mechanisms have influence on the transition temperature $T_c$ of mixed compounds. In the following this will be discussed more in detail.

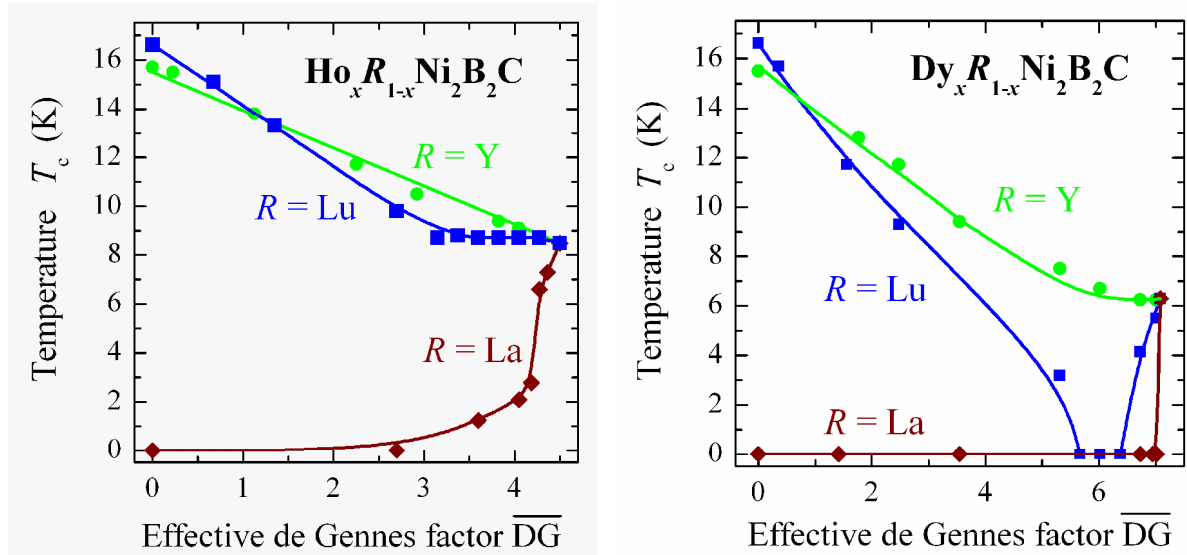

**Figure 1.** Dependence of the superconducting transition temperature $T_c$ of $Ho_xR_{1-x}Ni_2B_2C$ (left) and $Dy_xR_{1-x}Ni_2B_2C$ (right) on the effective de Gennes factor $\overline{DG}$ for $R$ = Y, Lu and La. The symbols at zero temperature mark values of $\overline{DG}$ where no superconductivity has been observed above 2 K.

*3.1. Nonmagnetic borocarbides*

Already in the very early studies on the effect of impurities in superconductors on their superconducting transition temperature $T_c$, Matthias et al. [2] showed that 1% of nonmagnetic Luthetium in the nonmagnetic superconductor Lanthanum reduces the superconducting transition temperature by about $\delta T_c = 0.1$ K as shown in Table 1.

In rare-earth nickel borocarbides magnetic moments only results from the 4$f$ electrons of the rare-earth elements whereas the Nickel does not contribute to localized or itinerant magnetic moments in these compounds [21]. Among the $R$Ni$_2$B$_2$C compounds with the nonmagnetic elements $R$ = Sc, Y, La and Lu only those with $R$ = Sc, Y, and Lu are superconductors whereas LaNi$_2$B$_2$C does not exhibit superconductivity down to $T = 300$ mK.

In Fig. 2 the superconducting transition temperatures $T_c$ of the Lu$_x$Y$_{1-x}$Ni$_2$B$_2$C and Sc$_x$Y$_{1-x}$Ni$_2$B$_2$C samples are shown in dependence on their lattice parameter $a$. Starting at pure YNi$_2$B$_2$C the value of $T_c$ decreases with increasing content of Lu or Sc until $a$ corresponds to the intermediate concentrations $x \approx 0.5$ where a minimum of $T_c$ is found and for further decreasing the Y content $T_c$ increases again. It will be difficult to understand these non-monotonic curves in detail. As discussed in Sec. 1 various "non-magnetic" mechanisms affect the superconductivity of mixed compounds. Although the 4$f$ electrons of the $R$ and $R'$ atoms do practically not participate in the Fermi surface of $R_xR'_{1-x}$Ni$_2$B$_2$C some variation of the itinerant-electron structure is expected due to the local chemical pressure resulting from the different sizes of the $R^{3+}$ and $R'^{3+}$ ions, being manifest also in the different sizes of the lattice constants of $R$Ni$_2$B$_2$C and $R'$Ni$_2$B$_2$C.

Thus, according to Lai et al. [22] the superconducting properties of $R$Ni$_2$B$_2$C and in particular $T_c$ are very sensitive to the Ni-Ni distance which is proportional to the lattice parameter $a$ because $a$ determines the in-plane Ni-Ni distance and, consequently, the electronic structure of the Ni(3d)-dominated conduction bands. This is the reason why $a$ is used as the abscissa in Fig. 2. The non-monotonic curves in Fig. 2 show that the properties of these mixed compounds can not be understood by only considering hypothetical homogeneous "gray systems" with averaged values of lattice constants and of other physical parameters. That is the fluctuations of these parameters have to be taken into account anyhow.

Furthermore, the scattering of conduction electrons off "impurities" i.e. deviations from the ideal LuNi$_2$B$_2$C-type lattice structure have to be considered, which is particularly important in these multiband superconductors [5].

A very simple phenomenological description of the effect of disorder on the superconducting properties has been proposed by Attfield et al. [6]. These authors quantified the reduction of $T_c$ of $R_2CuO_4$, caused by doping at the $R$ site by

$$T_c(x) = T_c^0(x) - p\sigma^2(x) \qquad (3)$$

where $p$ is some parameter characterising the strength of disorder and $T_c^0(x)$ is the dependence of $T_c(x)$ of the above mentioned gray system without disorder effects and $\sigma^2(x)$ is the variance of the ionic radii of the doping ions at the $R$ site. As can be seen in Fig. 2 this approach works surprisingly well also for the systems $Lu_xY_{1-x}Ni_2B_2C$ (red curve) and $Sc_xY_{1-x}Ni_2B_2C$ (blue curve).

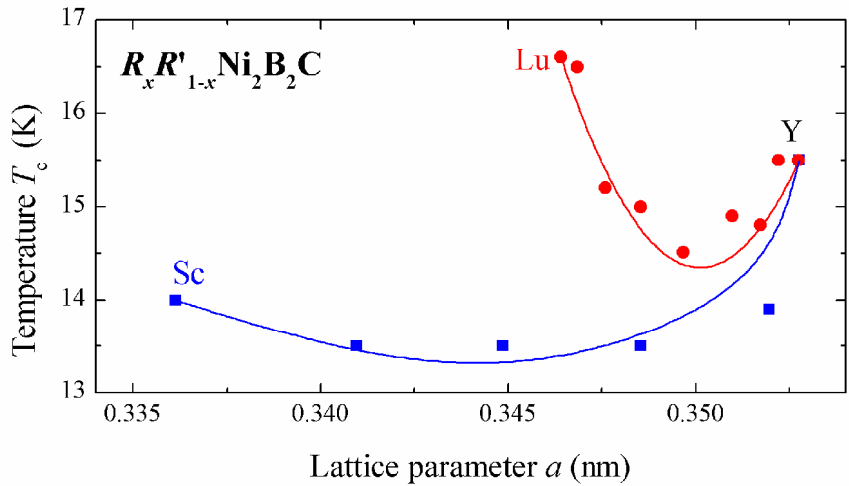

**Figure 2.** Superconducting transition temperature $T_C$ of Lu$_x$Y$_{1-x}$Ni$_2$B$_2$C (red circles) and Sc$_x$Y$_{1-x}$Ni$_2$B$_2$C (blue squares) compounds in dependence on their lattice parameters $a$. The red and blue lines are curves fitted to Eq. (3).

*3.2. Paramagnetic R impurities in the nonmagnetic $YNi_2B_2C$ superconductor*

According to Section 3.1, impurities introduced into the lattice of a superconducting material should always tend to reduce the superconducting transition temperature $T_c$, due to effects of local fluctuations of physical parameters such as the $R^{3+}$ ionic radii. If the impurities carry a magnetic moment at least one additional mechanism occurs, which further reduces $T_c$. This is well described by the Abrikosov Gor'kov theory (see Section 1.1) and its extensions for the presence of crystalline electric fields (CEF) acting on the $R^{3+}$ lattice sites [23] and for systems with higher impurity concentrations resulting in cooperative magnetic phenomena (see Ref. [5]). Using $^{57}Fe$ Mössbauer spectroscopy Sánchez et al. [24] have detected a pair-breaking field at the Ni site, caused by the $R$ magnetic moments. In the absence of CEF effects the suppression of $T_c$ by non-interacting paramagnetic impurities is given quantitatively by Eq. (1). This expression should well describe the suppression of $T_c$ of $Gd_xY_{1-x}Ni_2B_2C$ compounds which can be regarded as a convenient reference system for the study of pair-breaking by paramagnetic impurities in the family of $R_xY_{1-x}Ni_2B_2C$ compounds, since $Gd^{3+}$ is a spin-only ion i.e. it has a spherical charge density with no orbital momentum $L$ and, therefore, it is nearly insensitive to the CEF. Therefore and because $Gd^{3+}$ has the largest de Gennes factor $DG$ among the $R^{3+}$ ions Gd is the most effective magnetic pair breaker among the $R$ dopants. This is in accordance with the examples in Table 1. As expected, $Gd_xY_{1-x}Ni_2B_2C$ compounds show a dependence of $T_c$ on $\overline{DG}$ as predicted by Eq. (1), which can be seen in Fig. 3 where $T_c$ vs. $\overline{DG}$ curves are shown for $R_xY_{1-x}Ni_2B_2C$ compounds with $R$ = Gd, Ho, Dy, Er, Tb, Pr and Sm. The curves for the heavy rare earths $R$ = Ho, Dy, Er, Tb show a smaller depression of $T_c$ with increasing $\overline{DG}$ compared to the ideal curve for $R$ = Gd. This can be explained by the presence of CEF that reduce the magnetic degrees of freedom of these $R^{3+}$ ions, i.e. the available space for fluctuations and scattering of their local magnetic moments (see Refs. [25,26, 5] as predicted in Ref. [23]. The curves for $R$ = Pr and Sm in Fig. 3 are much steeper than that for $R$ = Gd and they can clearly not be described by the Abrikosov Gor'kov formula (Eq. 1). This is attributed to the much larger ionic radii of $Pr^{3+}$ and $Sm^{3+}$.

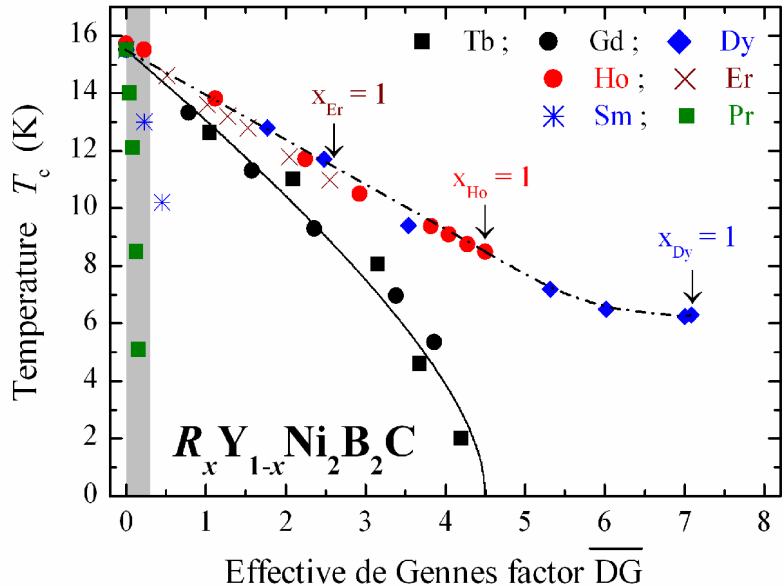

**Figure 3.** Superconducting transition temperature $T_c$ of $R_xY_{1-x}Ni_2B_2C$ compounds in dependence on the effective de Gennes factor $\overline{DG}$ for various elements $R$. The thick line corresponds to Eq. (1). Chain dotted line: guide for the eye for $R$ = Er, Ho and Dy.

In the following we will study the combined effects of the deviation $\delta r$ of the ionic radii of the impurity ions from the radius of $Y^{3+}$ and of the Abrikosov Gor'kov-type magnetic pair breaking on various $R_{0.05}Y_{0.95}Ni_2B_2C$ compounds (see Fig. 4). For such diluted mixed compounds Eq. (1) can be linearized with respect to the impurity concentration $x$. This results in

$$\delta T_c \propto N(E_F) I^2 \overline{DG}, \qquad (4)$$

where $\delta T_c$ is the difference between the transition temperature $T_c$ at impurity concentration $x$ and that at zero concentration. In order to visualise the combined influence of the different pair-breaking mechanisms, in Fig. 4 both are shown, the values of $\delta T_c$ and the depression rates $\delta T_c / \overline{DG}$. Taking into account the results of Sec. 3.1, in particular Eq. (3), a parabolic fit of the $\delta T_c / \overline{DG}$ values is shown in Fig. 4, too. With the exception of $R$ = Tm the deviations of $\delta T_c / \overline{DG}$ from the depicted parabolic fit are within the experimental errors of the $\delta T_c$ values. The enhanced depression rate for $R$ = Tm for which $|\delta r|$ has the same value as for $R$ = Gd seems to be connected with an enlarged exchange interaction parameter $I$ of this magnetically-easy-$c$-axis material [27, 28]. It can be clearly seen in Fig. 4 that for the heavy rare-earth elements $R$ = Er, Ho, Dy, Tb and Gd the depression rate $\delta T_c / \overline{DG}$ is almost independent of $\delta r$ whereas for light rare-earths $R$ = Sm, Nd and Pr, $\delta T_c / \overline{DG}$ increases strongly with increasing $\delta r$. Thus in $R_xY_{1-x}Ni_2B_2C$ compounds with light rare-earth elements $R$ disorder effects are dominating in the reduction of the superconducting transition temperature, whereas in cases of heavy rare earths $R$ magnetic pair-breaking dominates.

It should be noted that the larger depression rate $\delta T_c / \overline{DG}$ observed in $Ho_xLu_{1-x}Ni_2B_2C$ to that in $Ho_xY_{1-x}Ni_2B_2C$ (at small $x$ or small $\overline{DG}$ in Fig. 1) can also be explained by the larger difference of the ionic radii of $Ho^{3+}$ and $Lu^{3+}$ compared of that of $Ho^{3+}$ and $Y^{3+}$, which results in larger contributions to the depression rate as described in Section 3.1.

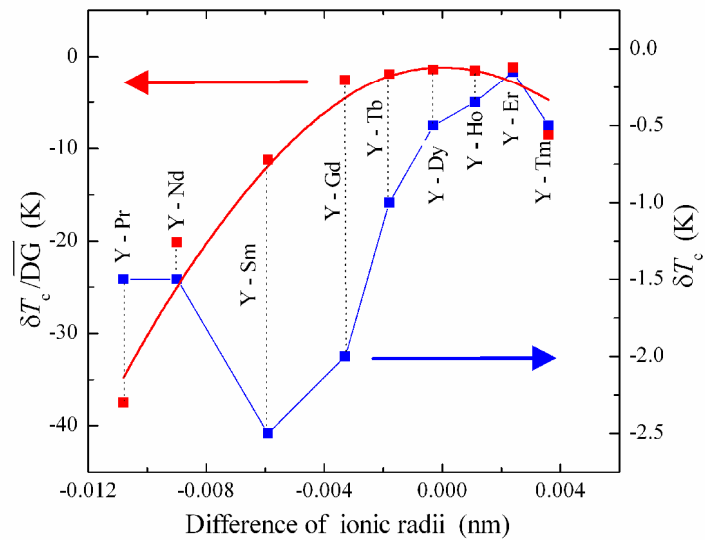

**Figure 4.** Reduction $\delta T_c$ of the superconducting transition temperature $T_c$ (blue symbols) and depression rate $\delta T_c / \overline{DG}$ (red symbols) for various $R_{0.95}Y_{0.05}Ni_2B_2C$ compounds in dependence on the difference $\delta r$ in the ionic radii of $Y^{3+}$ and $R^{3+}$. The red line is a parabolic fit according to Eq. (3) and the blue line is a guide for the eyes.

*3.3. Non-magnetic R' impurities in antiferromagnetic $RNi_2B_2C$ superconductors*

It can be seen in Fig. 1 that small concentrations of the non-magnetic elements $R'$ = Y, Lu or La in $R'_x R_{1-x} Ni_2 B_2 C$ superconductors with $R$ = Ho or Dy result in a large variety of different behaviours where the most surprising difference is that between $Lu_x Ho_{1-x} Ni_2 B_2 C$ and $Lu_x Dy_{1-x} Ni_2 B_2 C$. The reason for this observation is that, in the considered ranges of concentrations $x$ and superconducting transition temperatures $T_c$, the $Lu_x Ho_{1-x} Ni_2 B_2 C$ compounds are paramagnetic i.e. they are above their magnetic ordering temperature whereas $Lu_x Dy_{1-x} Ni_2 B_2 C$ is a true antiferromagnetic superconductor. It should be noted that in the considered range of small Lu concentrations $x$ the transition temperature $T_c$ of $Lu_x Dy_{1-x} Ni_2 B_2 C$ even increases with increasing effective de Gennes factor $\overline{DG}$, which seemingly is in contradiction to the results of Section 3.2. The corresponding steep branch of the curve in Fig. 1 for $Lu_x Dy_{1-x} Ni_2 B_2 C$ at small $x$ can be interpreted as being based on electron scattering on non-magnetic Lu impurities in the antiferromagnetic superconductor $DyNi_2B_2C$. This strong depression of superconductivity has been interpreted as pair breaking due to creation of magnetic holes [29]. However as discussed in Section 1.3 it had been shown in earlier theoretical analyses [7,8] that other effects of non-magnetic impurities should also be efficient in suppression of superconductivity (see also Ref. [10]). As reported and interpreted in Section 1.3 the results published on Dy-rich compounds $Y_x Dy_{1-x} Ni_2 B_2 C$ exhibit much scatter [30,31,32]. Therefore this system will not be discussed further in this study.

The small but interesting difference in the curves for $Y_x Ho_{1-x} Ni_2 B_2 C$ and $Lu_x Ho_{1-x} Ni_2 B_2 C$ at small $x$ in Fig. 1 is not yet understood and has to be interpreted in the scenario of multiband superconductivity in these quaternary borocarbides (see Ref. [5]).

In both systems $La_x Ho_{1-x} Ni_2 B_2 C$ and $La_x Dy_{1-x} Ni_2 B_2 C$ there is a very strong decrease of $T_c$ if $x$ increases starting at $x = 0$. In $La_x Ho_{1-x} Ni_2 B_2 C$ diluting Ho by La the large difference in the ionic radii of $Ho^{3+}$ and $La^{3+}$ leads to a certain decrease of $T_c$ with decreasing Ho content until $T_c$ reaches the antiferromagnetic ordering temperature. Then $T_c$ decreases much steeper with increasing $x$ (see Fig. 1), because now the non-magnetic La impurities are hosted by a superconducting antiferromagnet, similar as in the case of $La_x Dy_{1-x} Ni_2 B_2 C$ [17,33].

In Fig. 5, the reduction $\delta T_c$ of the superconducting transition temperature in $Ho_{0.99} R'_{0.01} Ni_2 B_2 C$

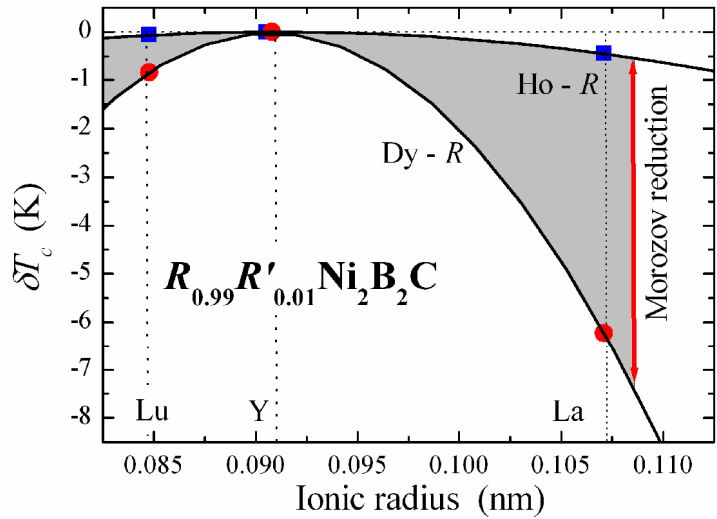

**Figure 5.** Reduction $\delta T_c$ of the superconducting transition temperature of $Ho_{0.99} R'_{0.01} Ni_2 B_2 C$ and $Dy_{0.99} R'_{0.01} Ni_2 B_2 C$ compounds with $R'$ = Y, Lu, La in dependence on the difference $\delta r$ in the ionic radii of $R'^{3+}$ and $Ho^{3+}$ and $Dy^{3+}$, respectively. The lines are guides for the eyes. The arrow "Morozov reduction" marks the difference between the paramagnetic Ho based and antiferromagnetic Dy based compounds.

and Dy$_{0.99}$R'$_{0.01}$Ni$_2$B$_2$C compounds with R' = Y, Lu, La is compared. Because Ho$_{0.99}$R'$_{0.01}$Ni$_2$B$_2$C is in the paramagnetic state, whereas Dy$_{0.99}$R'$_{0.01}$Ni$_2$B$_2$C is antiferromagnetically ordered, this comparison allows to separate the two contributions to $\delta T_c$ in the Dy based compounds - the contribution due to differences in ionic radii from that due to pair breaking of non-magnetic impurities in the antiferromagnetic superconductor. The latter one (termed "Morozov reduction") corresponds to the difference of both curves in Fig. 5.

## 4. Conclusions

The investigation of pseudoquaternary compounds (R,R')Ni$_2$B$_2$C with two different rare earths on the rare earth lattice site revealed insight into the pair breaking mechanisms in mixed-compounds superconductors. Three mechanisms concerning the reduction of $T_c$ could be distinguished. (i) Disorder always modifies the superconducting state und in particular the value of $T_c$. This is no "pair breaking" and, therefore it is not in contradiction to Andersons theorem. For the here investigated materials the disorder effect on $T_c$ could be successfully scaled with the variance of the $R^{3+}$ ionic radii. (ii) Paramagnetic R' impurities in a nonmagnetic RNi$_2$B$_2$C superconductor cause Abrikosov-Gor'kov type magnetic pair breaking combined with the mentioned disorder effect. (iii) If the RNi$_2$B$_2$C host superconductor is an antiferromagnet that contains non-magnetic R' impurities the disorder effects are combined with the pair breaking effect described by Morozov. It would be interesting to analyze also the influence of R' = Ce or Yb impurities in RNi$_2$B$_2$C superconductors because in these elements the 4f electrons stronger tend to hybridize with the conduction electrons compared to the behavior of the other R' elements.

## Acknowledgements

In writing this article the authors had close cooperation with S.-L. Drechsler, A. Kreyssig, A. Handstein, M. Loewenhaupt and H. Rosner. This work has been supported by DFG (SFB463) and IFW Dresden.